\documentclass{article}
\usepackage{arxiv}
\usepackage{xcolor}
\usepackage{amssymb}

\title{Photoluminescent colour centres on a mainstream silicon photonic foundry platform}
\author{Prosper Dellah Allo \and A. Aadhi \and Amirhossein Mosaddegh Yengejeh \and Hazel Bakajsa \and Mirabel N. M. Mensah \and Marcus Tamura \and Bhavin J. Shastri \and Alexander N. Tait\\
Department of Electrical and Computer Engineering, Queen's University, Kingston, ON K7L 3N6 Canada\\\\ Centre for Nanophotonics, Department of Physics, Engineering Physics \& Astronomy, \\Queen's University, Kingston, ON K7L 3N6, Canada.}

\usepackage{lineno} 
\usepackage{graphicx} 
\usepackage{amsmath} 
\usepackage{hyperref} 
\usepackage[final]{fixme} 

\begin{document} 
\maketitle

\begin{abstract}
    The fabrication of silicon photonic components in commercial CMOS-compatible foundries has revolutionized the impact of silicon photonics on advancing communication, quantum computing and artificial intelligence, due to their benefits of mass production, high throughput, low cost, and high performance.
    The indirect bandgap of silicon introduces a fundamental challenge; thus, the mainstream silicon-on-insulator (SOI) platform does not have efficient light sources. Recently, luminescent colour centres in SOI have emerged as one promising approach for developing efficient on-chip classical and quantum light sources, although past work has relied on custom fabrication that is not foundry-compatible.
    In this work, we demonstrate W-centre photoluminescence on a mainstream silicon photonics platform through development of a straightforward back end-of-line (BEOL) treatment. At an optimal implant energy of 7~MeV, we observed W-centre photoluminescence with a brightness comparable to prior in-house processes. We performed a series of experiments on Circular Bragg Grating (CBG) devices with varying pitches, duty cycles, and implant energies to confirm the PL emission from the encapsulated SOI device layer rather than the handle wafer.
    Our novel approach in fabricating silicon colour centres in commercial silicon photonic foundry processes opens up new opportunities for integrating classical and quantum light sources directly onto silicon photonic circuits, unlocking opportunities for large-scale integration of advanced photonic architectures on chip.
\end{abstract} 


\section{Introduction} \label{sec:introduction}
    Silicon photonics has emerged as a transformative technology with potentials to impact the growing energy consumption in data centres, communication networks, and computing infrastructure. It has also found applications in biomedical applications, computation, and spectroscopy~\cite{zhou2023prospects, han2022recent, zhou2015lowering}. The platform provides a balance cost-effectiveness, potential to scale, and a highly-functional device library~\cite{lim2013review, xiang2021perspective}.
    These advantages are made possible by leveraging the existing foundry infrastructure developed for the microelectronics industry.

    The economic equation of the standard platform is unlike any 20$^\text{th}$-century approach for fabricating PICs; however, there is one key component lacking.
    Since the advent of silicon photonic integrated circuit (PICs), there has been a search for an efficient and simple light source~\cite{wu2014advancing}. Due to its indirect bandgap, the radiative efficiency of crystalline silicon(Si) is low ($\eta_{int}\sim10^{-6}$~\cite{iyer1993light, fiory2003light}).
    Without an integrated light source, there is a need for delicate, one-off fibre coupling procedures. While these have improved immensely, they still impose harsh input/output (I/O) limitations.

    Various approaches have been developed to circumvent the indirect bandgap of silicon in order to achieve light emission within a silicon PIC.
    These approaches include (among others) heterogeneous and/or hybrid integration of III-V materials
    ~\cite{liang2010recent, Roelkens:10, Liao:18, adomLi:22},
     doping with rare earth elements~\cite{Belt:14}, Raman effect devices with off-chip pumping~\cite{rong2005all}, and strain engineering of germanium~\cite{Bao:17}. Each approach has complementary advantages and drawbacks. While specialized III-V/SOI platforms are commercially available, the cost of these is at least an order-of-magnitude higher than a mainstream actives process. The high cost of integrating III-V sources neutralizes several of the key advantages of silicon photonics in the first place such as 300~mm wafer-scale production, high yield, and/or chemical compatibility with silicon-based facilities sensitive to cross-contamination.

\begin{figure}[htbp]
        \centering
        \includegraphics[width=1.0\textwidth]{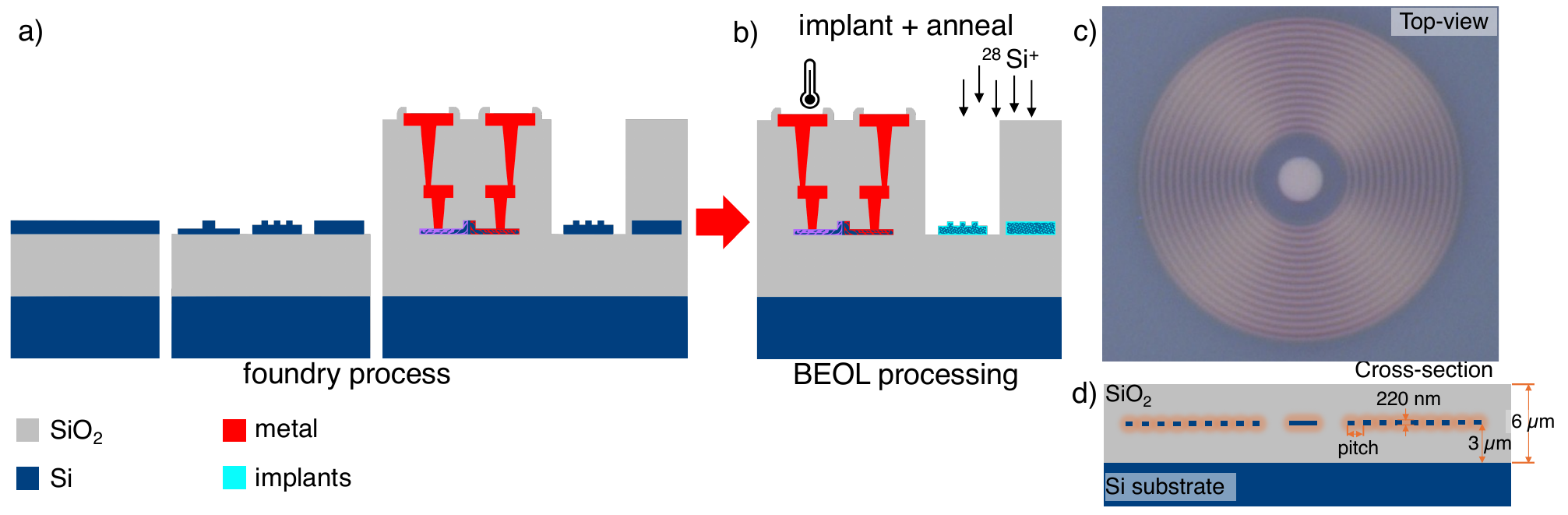} 
        \caption{\textbf{Device fabrication process} (a) Foundry fabrication process. (b) Implantation and annealing process. (c) Microscopic image of Circular Bragg Grating (CBG). (d) Cross-section description of CBG.} 
        \label{fig:foundry_implant_image} 
    \end{figure}

    One particular platform for active modulators and detectors has emerged as a standard that is offered by multiple foundries around the world~\cite{chrostowski2019silicon}. We refer to this platform as the ``mainstream'' or standard.
    This standard process is inexpensive enough for classroom teaching and readily scalable to the needs of small business and industrial production~\cite{darcie2021siepicfab}.
    While some front-end of line (FEOL) variants of this standard process are offered commercially, reliance on non-standard processes counteracts several of the key advantages that make silicon photonics attractive in the first place.
    Deviations from this mainstream platform involve changes inside the foundry, a.k.a. front-end of line (FEOL) process changes.
    Figure~\ref{fig:foundry_implant_image}a shows a standard foundry process, which features
    3-layer SOI etches with sub-wavelength resolution,
    6-layer dopants for modulators,
    epitaxial germanium for photodetectors,
    2-layer routing metals and vias.
    These PICs are delivered with a 3~$\mu$m of $SiO_2$ encapsulation layer.

    The prospect of all-silicon light sources would greatly enhance its potential to leverage manufacturing economies of scale~\cite{shainline2007silicon}.
    Defect centres in silicon offer promising solutions for wafer-scale integration of on-chip light sources that are more compatible with the mainstream platform.
    There has been a recent acceleration of research on colour centres on SOI platforms. These include W-centres~\cite{ buckley2020optimization, tait2020microring, kizhake2024enhanced}, T-centres~\cite{macquarrie2021generatingTcentres}, and G-Centres~\cite{day2024electrical}.
    These defect centres are formed through the controlled addition of lattice damage sites, which are then annealed to form emissive defects.
    Defect centres create electron states within the silicon bandgap.
    These discrete states support efficient radiative transition in a bright zero-phonon line (ZPL) and a phonon sideband.

    Defect centres on SOI platforms has been previously studied using custom in-house fabrication processes~\cite{ prabhu2023individually, islam2023cavity, jhuria2024programmable}. They are a step closer to compatibility with wafer-scale integration~\cite{buckley2020optimization}, but not quite compatible with the standard platform.
    Active foundry PICs must arrive with an oxide encapsulation if they are to contain the active devices. This creates a nontrivial process incompatibility because prior studies require the defect implantation to occur prior to oxide encapsulation and metallisation.
    If these processes were to be mapped to a foundry platform, they would face similar challenges as other light source candidates, namely, FEOL process modification that neutralize several of the advantages of using the silicon photonics platform. To the authors' knowledge, silicon photoluminescence has not been observed on a PIC that was delivered from a mainstream foundry process.

    In this work, we demonstrate W-centre photoluminescence (PL) on PICs fabricated on a mainstream silicon foundry platform.
    We address the critical challenge of creating W centres through a 3~$\mu$m encapsulation oxide through development of a high-energy, back-end-of-line (BEOL) recipe for ion implantation and W-centre formation.
    The observed PL intensity is comparable to that of an unencapsulated reference sample that reproduces the W-centre recipes used in prior work on SOI~\cite{kizhake2024enhanced}.
    We provide several forms of evidence to support the claim that the PL emission comes from the SOI layer, rather than the handle wafer.
    The W centre is favourable for research because of its high brightness and relatively simple fabrication process.
    That said, this demonstration is versatile and potentially adaptable to other defect centres, including T and G centres, using similar fabrication processes and implantation methods as employed in this work.
    As a result, our demonstration is a significant step towards mainstream foundry-based on-chip classical and quantum light sources, with potential applications in communications, computing, and quantum technology.

\section{Methods} \label{sec:methods}

    \subsection{Device Fabrication} \label{sub:device_fabrication}
        Samples were fabricated using the Advanced Micro Foundry (AMF) active silicon photonic process (AMFSiP).
        The foundry fabrication process is depicted in Figure~\ref{fig:foundry_implant_image}a.
        The SOI wafer has a 220~nm device layer on a 3~$\mu$m buried oxide (BOX) on a 600~$\mu$m handle wafer. First, the foundry etches the device silicon. Next, it implants dopants for modulators and grows Ge for detectors. Next, it performs two iterations of oxide deposition, vias, and wiring. Finally, it opens the top oxide according to the user design. While the oxide cannot be opened over active devices, we open some passive devices to use as a reference.
        The SOI layer arrives buried under a 3~$\mu$m cladding oxide.

        Figure~\ref{fig:foundry_implant_image}b illustrates the BEOL W-centre formation process performed in a university cleanroom (NanoFab Kingston).
        Si ions ($^{28}$Si$^+$) were blanket-implanted at a 7$^\circ$ tilt angle to avoid ion channelling~\cite{takeda1986precise}.
        During implantation, atoms in the silicon lattice are displaced; therefore, the atom and recoiling ions settle as interstitials, leaving vacancies and interstitial damage along their paths.

        After irradiation, the implanted photonic chips are annealed at 250$^\circ$C for 30~minutes in an air environment using a modified household air fryer (CRUX Indoor Grill \& Air Fryer 9-Qt \textit{17176}). This annealing process allows Si damage sites to migrate and form clusters of silicon interstitials (n = 3, 4, 5, etc.)~\cite{ takeda1986precise, johnston2024cavity}. 
        The luminescent W centre is an n = 3 tri-interstitial cluster, while n = 4, n = 5, and higher clusters are absorbing. The intensity of the PL depends on a ratio between the rate of the radiative recombination at the W centre and the rates of other non-radiative processes at other defects.
        Anneal parameters were chosen to match those found to be optimal by prior work~\cite{buckley2020optimization}.
        This enables efficient radiative recombination involving localized states, producing photoluminescence at 1218~nm (1.018~eV).

        We use two sets of Si$^+$ ion implant parameters (energy and fluence). A reference sample without a cladding oxide layer was blanket-implanted at 60~keV energy with a fluence of $5 \times 10^{12}$~at/cm$^2$ in order to reproduce the recipe presented in previous work~\cite{buckley2020optimization}. 

        Encapsulated samples require significantly higher energies and fluences.
        We used the Stopping Range of Ions in Matter (SRIM) simulation\cite{ziegler1995SRIM} to estimate of implantation energy and fluence range that would be needed to form W centres underneath the 3~$\mu$m thick cladding oxide of the foundry-fabricated chips. According to simulations, 5~MeV would optimize Si$^+$ ion stopping depth, although stopping range is not the same as defect formation depth. Prior work has shown that ion stopping range overestimates energies that yield optimal formation of W centres, so we chose an energy parameter range covering (4.4-9.0~MeV).
        All encapsulated samples were implanted with fluence of ($5\times10^{13}$~at/cm$^{-2}$), 10x more the unencapsulated reference. This fluence was chosen in anticipation that the thick oxide would reduce the frequency of damage site formation. The risk of overshooting fluence was considered acceptable because PL brightness did not exhibit significant quenching with fluence in~\cite{buckley2020optimization}.

        Three types of photonic structures are used in experiments. A large 0.9 x 0.4~mm SOI rectangle is referred to below as unpatterned. An area with no device layer is used to measure only the handle wafer. Patterned circular Bragg grating (CBG) structures, a.k.a. bullseye, are designed
        to nominally diffract in-plane PL upwards towards the objective to improve collection efficiency~\cite{ kizhake2024enhanced}. We fabricated CBGs with varying design parameters (pitch and duty cycle) to optimize PL collection enhancement. The microscopic image of a W-centre implanted CBG and its cross-section is shown in Figure~\ref{fig:foundry_implant_image}c.
        The CBGs used in this study were designed by varying the pitch and duty cycle according to the relation: $\text{rib width} = \text{duty cycle} \times \text{pitch}$.

    \subsection{Experimental Setup} \label{sub:experimental_setup}

        \begin{figure}[htbp]
            \centering\includegraphics[width=0.5\textwidth]{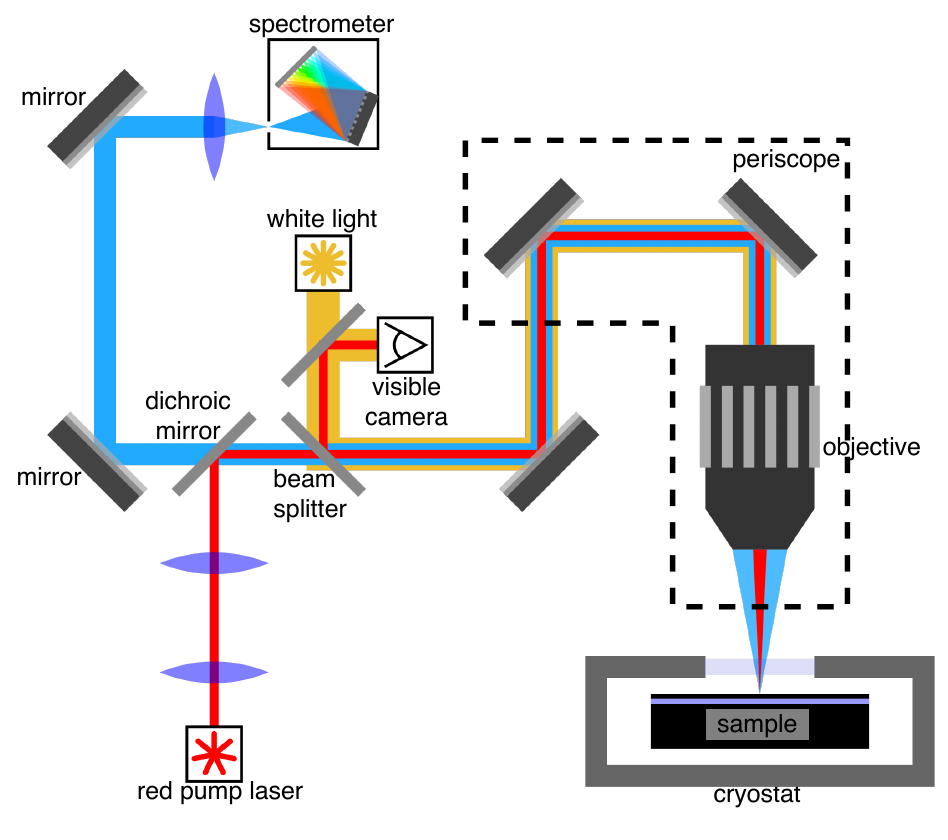}
            \caption{\textbf{Experimental setup.} It consists of a 635~nm pump laser, a cryo-cooled spectrometer, a 4K cryostat, a periscope, infinity corrected objective lens, a telescope, and a visible camera.}
            \label{fig:experimental_setup}
        \end{figure}

        Figure 2 depicts the cryogenic micro-photoluminescence ($\mu$PL) experimental setup. The sample was mounted on a copper mount attached to the cold stage of an optical access 4~Kelvin cryostat, equipped with vibration isolation bellows (ARS DMX-20).
        In all experiments below, the sample is held at 6~K since it is more practical to stabilize at 6~K than at 4~K.
        A continuous-wave (CW) HeNe laser at 635~nm was used to optically pump the W centres in the mounted sample through a quartz window. A pair of lenses with focal lengths of 50~mm and 75~mm were arranged as a telescope to adjust the pump beam spot size on the sample. An infinity-corrected  objective
        Mitutoyo MY10X-823 with an 0.26~NA, 10x magnification, and an 200~mm working distance was mounted on optomechanical stages in a periscope configuration, allowing XYZ movement.

        The pump beam was directed and focused through the objective onto a target device using XYZ motorized translation control. The optically pumped device emits PL in all directions. PL emitted normal to the surface is collected by the same objective lens. A dichroic mirror was used to separate the back reflected pump beam from the measured PL. The spectrometer (Andor Kymera 328i), equipped with a 512-pixel linear InGaAs detector array (iDus 490), measures the photon counts from visible to infrared wavelengths. A broadband white light source and a visible camera were used to image, identify, and align to devices on the sample.

\section{Results} \label{sec:results}
    \subsection{Unencapsulated reproduction} \label{sub:validation}
        \begin{figure}[htbp]
            \centering
            \includegraphics[width=1.0\textwidth]{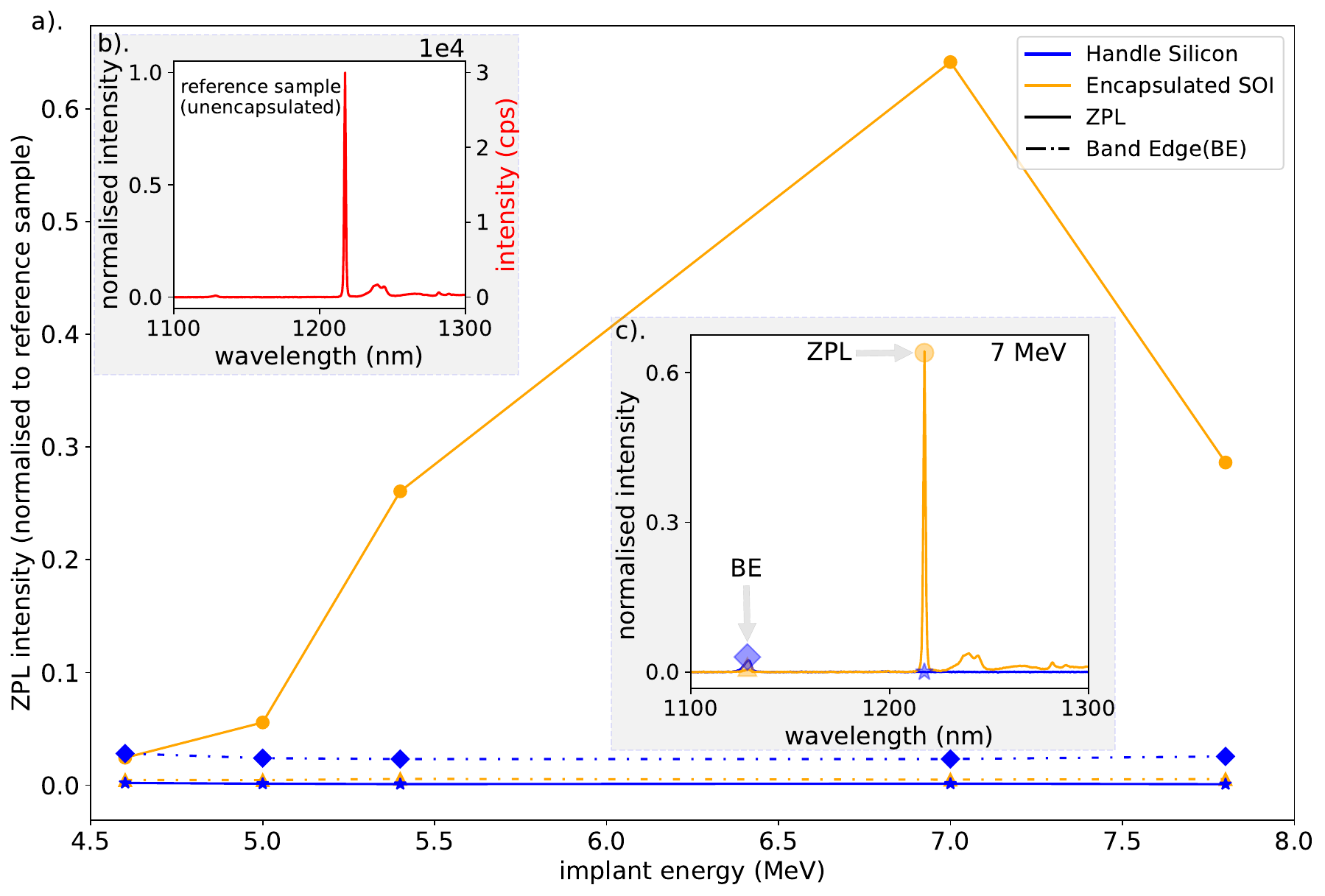}
            \caption{ \textbf{Emission spectra based on implant energies.}
            a. Emission spectrum from the reference SOI device of the unencapsulated sample used as reference in this study.
            b. Emission spectrum of encapsulated sample implanted at 7~MeV.
            c. Comparison of peak-ZPL-intensity
            (\textcolor{orange}{$\bullet$} and \textcolor{blue}{$\bigstar$})
            and peak-band-edge-intensity
            (\textcolor{blue}{$\blacklozenge$} and \textcolor{orange}{$\blacktriangle$}) from the handle silicon(blue) and the encapsulated device layer(orange) against the implant energy of the implanted samples.}
            \label{fig:energies_spectra}
        \end{figure}

        We demonstrate our mainstream-compatible BEOL fabrication process first using an unencapsulated reference sample that attempts to reproduce the fabrication parameters of~\cite{buckley2020optimization}, i.e. 60~keV energy with a fluence of $5 \times 10^{12}$~at/cm$^2$, annealed at 250$^\circ$C for 30~minutes.
        Our fabrication methods differ from that work in terms of PIC production facility, ion implantation facility, and annealing atmosphere.
        The $\mu$PL spectrum of the reference sample is shown in Fig.~\ref{fig:energies_spectra}b (inset). As expected, the spectrum reproduces a W-centre profile with zero-phonon line (ZPL) at 1218~nm and a phonon sideband spanning 1225~nm–1275~nm.
        We observed a similar temperature dependence: PL brightness has a sharp turn at 40~K, below which it varies by less than 20\%.

        The reference ZPL brightness was measured to be $3.0 \times 10^4$ counts per second at 90~$\mu$W of pump power (330cps/$\mu$W).
        This unencapsulated reference sample was used as a PL brightness standard to correct for any variation in collection efficiency that can occur between different cooldowns. All plots referring to "normalized intensity" below define unity as the brightness of this reference sample.

        This reproduction is significant in terms of demonstrating W-centre PL on a foundry PIC with BEOL processing; however, its impact is limited in and of itself. Without encapsulation, this recipe would only be applicable to passive foundry devices.

    \subsection{Encapsulated sample recipe} \label{sub:new_recipe}
        Figure~\ref{fig:energies_spectra}a shows the PL from encapsulated foundry samples implanted at different energies: 4.6, 5.0, 5.4, 7.0, and 7.8~MeV, each implanted at a fluence of $5 \times 10^{13}$~at/cm$^2$. All ZPL intensities were normalized with respect to the reference sample to quantitatively compare the PL emissions across all devices.
        Of these samples, a maximum occurred at 7.0~MeV -- two orders-of-magnitude higher implant energy than unencapsulated SOI optimum. Its brightness was 64\% of the reference, thus confirming the ability to induce W-centre formation at approximately the same efficiency, even through a 3~$\mu$m encapsulation oxide. It appears likely that the true optimum lies somewhere between the 5.5~MeV and 7.0~MeV data points and could be refined with a denser parameter sweep.

        The high implant energies introduce a significant possibility that W centres also formed in the handle wafer underneath the BOX.
        To confirm that the observed PL emission is from the SOI device layer and not the handle, PL spectra were measured at two locations on each sample: (1) with unpatterned 220~nm device SOI (orange) and (2) with device SOI fully etched away (blue). We used band-edge emission at 1120~nm as a way to establish a focus when measuring the bulk handle.
        The inset Fig.~\ref{fig:energies_spectra}c compares the spectra of device layer vs. bulk handle measurements. There was no detectable ZPL from bulk handle despite the bulk showing brighter band edge. This finding was true at all energies tested, which can be seen in Fig.~\ref{fig:energies_spectra}a dashed lines corresponding to band-edge intensity.
        Both the ZPL and band-edge intensities show a linear relationship with pump power and remain well below the saturation intensity.
        These measurements confirm the formation of W-centres happened within the photonic device layer underneath the 3~$\mu$m cladding oxide.

    \subsection{Micro-PL in passive devices} \label{sub:bullseyes}
        \begin{figure}[htbp]
            \centering\includegraphics[width=1.0\textwidth]{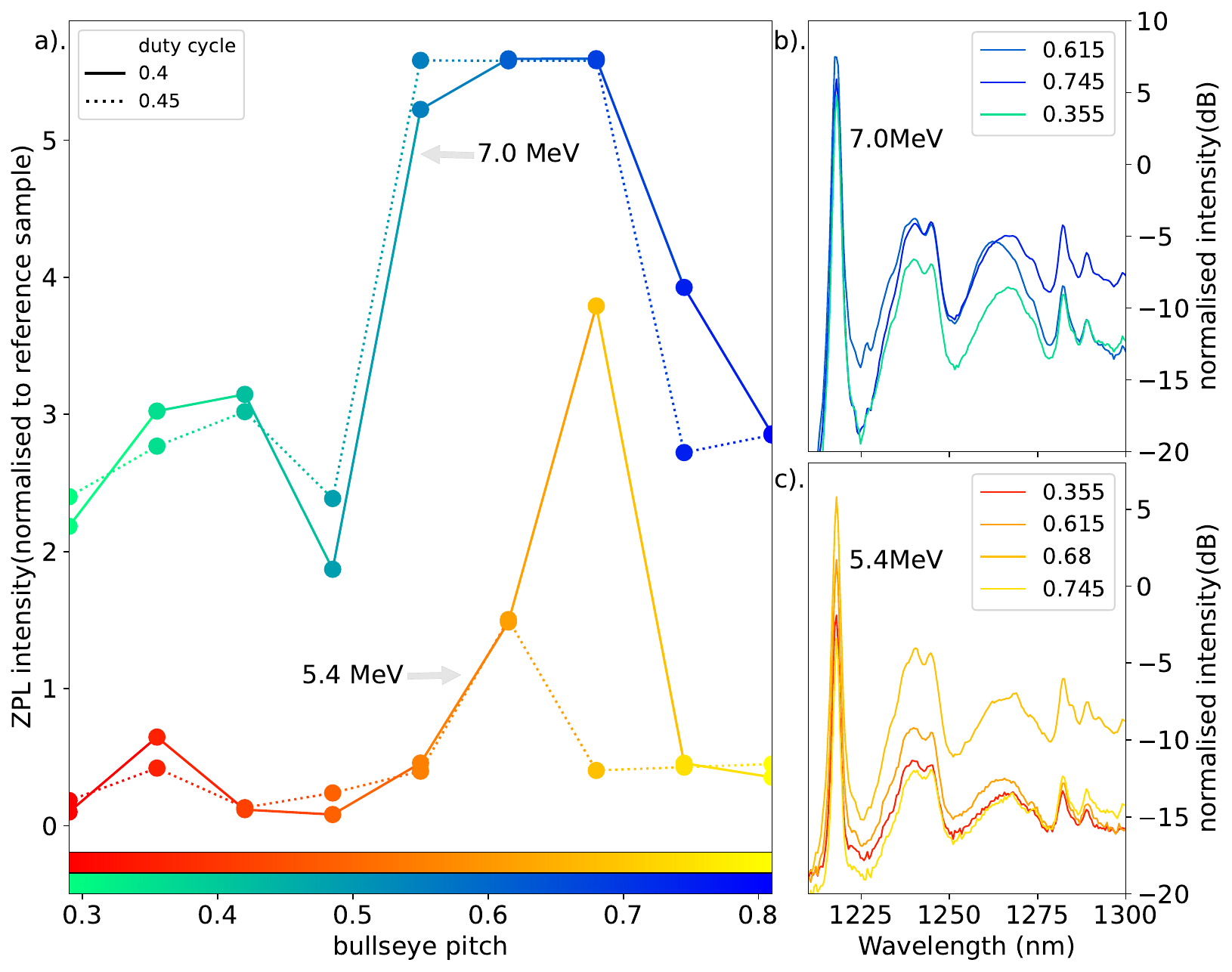}
            \caption{ \textbf{Emission spectra based on bullseye pitch and duty cycle.}
            a). Comparison of peak-ZPL-intensity from CBGs with varying pitches and duty cycles, implanted at 5.4~MeV(red to yellow) and 7~MeV(green to blue).
            b). Emission spectra of circular Bragg gratings implanted at 7~MeV.
            c). Emission spectra of circular Bragg gratings implanted at 5.4~MeV. 
                }
            \label{fig:pitch_spectra}
        \end{figure}

        To further confirm that W-centre PL originates from the encapsulated device layer, rather than the handle wafer, we measured patterned photonic structures. We designed various bullseye coupler devices with differing pitches and duty cycles (micrograph in Fig.~\ref{fig:foundry_implant_image}c,d), then compared them across dies with a range of implant energies. If the PL emission primarily originated from the handle wafer, it would exhibit constant intensity due to the uniform formation of W-centres in the handle silicon, regardless of variations in pitch or duty cycle.
        Figure~\ref{fig:pitch_spectra}a presents the ZPL peak intensity as a function of pitch for two different duty cycles (0.4 and 0.45) and two implant energies (5.4~MeV and 7~MeV). Dotted lines correspond to PL emission with a duty cycle of 0.45, while solid lines correspond to the same pitch and implant energies with a duty cycle of 0.4.

        As evident from Figure~\ref{fig:pitch_spectra}a, PL intensities vary strongly with pitch, duty cycle, and implant energy, confirming that the measured PL emission originates from the encapsulated SOI device layer. For a duty cycle of 0.4, the optimal CBG design has a pitch of 0.745~$\mu$m. For a duty cycle of 0.45, the optimized pitch is 0.68~$\mu$m.
        Figure~\ref{fig:pitch_spectra}b illustrates the variation in emission spectra for implant energies of 7~MeV (green to blue) and 5.4~MeV (red to yellow) across different pitches. The spectra reveal wavelength-dependent variations in collection efficiency in the phonon sideband, with certain wavelengths showing improved collection efficiency compared to others within the same spectrum.
        This observation also further validates that the pitch-dependent features in the emission spectra are exclusively due to the formation of W-centres in the encapsulated devices. Notably, this observation remains consistent at different implant energies.

\section{Conclusion} \label{sec:conclusion}
    We demonstrated W-centre formation and luminescence on samples delivered by a standard silicon photonic foundry. This demonstration was enabled by a straightforward BEOL recipe based on high-energy Si ion implantation.
    We performed a series of experiments to provide strong evidence that the emission originates from the SOI device layer and interacts with photonic passives designed on that layer.
    Using a reference sample with similar implant parameters as prior work on in-house-fabricated devices~\cite{buckley2020optimization}, we observed an intense zero-phonon-line emission.
    This result indicates that it is possible to maintain brightness even when implanting through the 3~$\mu$m encapsulation oxide that comes with foundry-fabricated samples.
    The existence of a broad phonon sideband will be an important diagnostic tool for cryogenic photonic device engineering, even without requirements for fibre coupling or wirebonds inside the cryostat.
    Further work into electrical pumping would enable fully-integrated cryogenic photonic systems.
    This approach opens several opportunities for further work on integrating bright light sources directly onto standardized silicon photonic platforms.

\section{Further Work}
    Colour centres offer a promising approach for realizing on-chip light sources for communication, signal processing, photonic computing and other applications. In particular, the CMOS-compatible colour centres demonstrated in this study, fabricated within a mature silicon photonic foundry, open up new possibilities for developing efficient light emitters using ultra-large-scale integration (ULSI) technology~\cite{PRXQuantum.5.010102, redjem2023all}.
    Our demonstration of photoluminescence (PL) from W-centres employs a novel approach to implantation and annealing conditions using foundry-fabricated chips. In addition, our demonstration of efficient PL emission from W-centres in foundry-fabricated Si devices, compatible with both optical and electronic functionalities in high-density silicon microelectronic circuits, facilitates large-scale electroluminescence, LEDs, and cryo-compatible integrations~\cite{ ebadollahi2024fabrication, buckley2017all}. W-centres have the potential to exhibit significantly enhanced cavity-coupled luminescence. Therefore, further integration of W-centres into microcavity designs could enable the development of efficient on-chip silicon lasers using foundry-fabricated emitters~\cite{tait2020microring}. Additionally, approaches such as focused-ion beam implantation~\cite{hollenbach2025programmable}, laser writing, and annealing techniques~\cite{quard2024femtosecond, jhuria2024programmable, gu2025end} offer promising pathways for creating programmable W-centres. These techniques facilitate efficient coupling of W-centres to microcavities, achieving high Purcell factors and efficient light emission. This includes the co-integration of single-photon nanowire detectors, cryogenic sensors, and quantum emitters, which further advance the development of silicon-based photonic technologies~\cite{PRXQuantum.5.010102}. Furthermore, our approach is applicable to other luminescent centres in silicon, such as T~centres and G~centres. However, additional studies are needed to optimize the implantation conditions and annealing processes for these centres. This work has significant implications for the fabrication of energy-efficient and scalable Si-based light sources and optical amplifiers using mainstream foundry techniques, paving the way for advancements in classical and quantum light sources, as well as other emerging applications, without the need for expensive and specialized fabrication processes.

\section*{Funding}
    This work was supported by the Canadian Foundation for Innovation (CFI) and Ontario Research Foundation (ORF) through the John R. Edwards Leaders Fund (JELF) program (Project 43419), the National Science and Engineering Research Council (NSERC) Discovery Grant (DG) program, and the Queen's University Summer Work Experience Program (SWEP). 

\section*{Acknowledgements}
    The authors gratefully acknowledge fabrication support from Graham Gibson at Nanofab Kingston (NFK), Kingston, Ontario, Canada.
    Ion implantation was performed by Martin Chicoine at the Canadian Charged Particle Accelerator Consortium (CCPAC), Montreal, Quebec, Canada.
    Foundry samples were fabricated by Advanced Micro Foundry (AMF) with multi-project wafer (MPW) support from the Canadian Microelectronics Corporation (CMC).
    Standard W-centre reference samples were provided by Jeff Chiles from the Quantum Nanophotonics and Faint Photonics groups at the National Institute of Standards and Technology (NIST), Boulder, CO, USA.

\section*{Disclosures}
    The authors declare no conflicts of interest.

\section*{Data availability}
    Data underlying the results presented in this paper are not publicly available at this time but may be obtained from the authors upon reasonable request.

\bibliography{references} 

\begin{thebibliography}{10}

\bibitem{zhou2023prospects}
Zhican Zhou, Xiangpeng Ou, Yuetong Fang, Emad Alkhazraji, Renjing Xu, Yating Wan, and John~E Bowers.
\newblock Prospects and applications of on-chip lasers.
\newblock {\em Elight}, 3(1):1, 2023.

\bibitem{han2022recent}
Yu~Han, Hyundai Park, John Bowers, and Kei~May Lau.
\newblock Recent advances in light sources on silicon.
\newblock {\em Advances in Optics and Photonics}, 14(3):404--454, 2022.

\bibitem{zhou2015lowering}
Zhiping Zhou, Bing Yin, Qingzhong Deng, Xinbai Li, and Jishi Cui.
\newblock Lowering the energy consumption in silicon photonic devices and systems.
\newblock {\em Photonics Research}, 3(5):B28--B46, 2015.

\bibitem{lim2013review}
Andy Eu-Jin Lim, Junfeng Song, Qing Fang, Chao Li, Xiaoguang Tu, Ning Duan, Kok~Kiong Chen, Roger Poh-Cher Tern, and Tsung-Yang Liow.
\newblock Review of silicon photonics foundry efforts.
\newblock {\em IEEE Journal of Selected Topics in Quantum Electronics}, 20(4):405--416, 2013.

\bibitem{xiang2021perspective}
Chao Xiang, Steven~M Bowers, Alexis Bjorlin, Robert Blum, and John~E Bowers.
\newblock Perspective on the future of silicon photonics and electronics.
\newblock {\em Applied Physics Letters}, 118(22), 2021.

\bibitem{wu2014advancing}
Shien-Yang Wu, CY~Lin, SH~Yang, JJ~Liaw, and JY~Cheng.
\newblock Advancing foundry technology with scaling and innovations.
\newblock In {\em Proceedings of Technical Program-2014 International Symposium on VLSI Technology, Systems and Application (VLSI-TSA)}, pages 1--3. IEEE, 2014.

\bibitem{iyer1993light}
Subramanian~S Iyer and Y-H Xie.
\newblock Light emission from silicon.
\newblock {\em Science}, 260(5104):40--46, 1993.

\bibitem{fiory2003light}
AT~Fiory and NM~Ravindra.
\newblock Light emission from silicon: Some perspectives and applications.
\newblock {\em Journal of Electronic Materials}, 32:1043--1051, 2003.

\bibitem{liang2010recent}
Di~Liang and John~E Bowers.
\newblock Recent progress in lasers on silicon.
\newblock {\em Nature photonics}, 4(8):511--517, 2010.

\bibitem{Roelkens:10}
G.~Roelkens, L.~Liu, D.~Liang, R.~Jones, A.~Fang, B.~Koch, and J.~Bowers.
\newblock Iii-v/silicon photonics for on-chip and intra-chip optical interconnects.
\newblock {\em Laser \& Photonics Reviews}, 4(6):751--779, 2019/11/04 2010.

\bibitem{Liao:18}
Mengya Liao, Siming Chen, Jae-Seong Park, Alwyn Seeds, and Huiyun Liu.
\newblock Iii--v quantum-dot lasers monolithically grown on silicon.
\newblock {\em Semiconductor Science and Technology}, 33(12):123002, 2018.

\bibitem{adomLi:22}
Nanxi Li, Guanyu Chen, Doris K.~T. Ng, Leh~Woon Lim, Jin Xue, Chong~Pei Ho, Yuan~Hsing Fu, and Lennon Y.~T. Lee.
\newblock Integrated lasers on silicon at communication wavelength: A progress review.
\newblock {\em Advanced Optical Materials}, 10(23):2201008, 2022.

\bibitem{Belt:14}
Michael Belt and Daniel~J. Blumenthal.
\newblock Erbium-doped waveguide dbr and dfb laser arrays integrated within an ultra-low-loss si3n4 platform.
\newblock {\em Opt. Express}, 22(9):10655--10660, May 2014.

\bibitem{rong2005all}
Haisheng Rong, Ansheng Liu, Richard Jones, Oded Cohen, Dani Hak, Remus Nicolaescu, Alexander Fang, and Mario Paniccia.
\newblock An all-silicon raman laser.
\newblock {\em Nature}, 433(7023):292--294, 2005.

\bibitem{Bao:17}
Shuyu Bao, Daeik Kim, Chibuzo Onwukaeme, Shashank Gupta, Krishna Saraswat, Kwang~Hong Lee, Yeji Kim, Dabin Min, Yongduck Jung, Haodong Qiu, Hong Wang, Eugene~A. Fitzgerald, Chuan~Seng Tan, and Donguk Nam.
\newblock Low-threshold optically pumped lasing in highly strained germanium nanowires.
\newblock {\em Nature Communications}, 8(1):1845, 2017.

\bibitem{chrostowski2019silicon}
Lukas Chrostowski, Hossam Shoman, Mustafa Hammood, Han Yun, Jaspreet Jhoja, Enxiao Luan, Stephen Lin, Ajay Mistry, Donald Witt, Nicolas~AF Jaeger, et~al.
\newblock Silicon photonic circuit design using rapid prototyping foundry process design kits.
\newblock {\em IEEE Journal of Selected Topics in Quantum Electronics}, 25(5):1--26, 2019.

\bibitem{darcie2021siepicfab}
Adam Darcie, Matthew Mitchell, Kashif Awan, Mahssa Abdolahi, Mustafa Hammood, Andreas Pfenning, Xiruo Yan, Abdelrahman Afifi, Donald Witt, Becky Lin, et~al.
\newblock Siepicfab: the canadian silicon photonics rapid-prototyping foundry for integrated optics and quantum computing.
\newblock In {\em Silicon Photonics XVI}, volume 11691, pages 31--50. SPIE, 2021.

\bibitem{shainline2007silicon}
Jeffrey~M Shainline and Jimmy Xu.
\newblock Silicon as an emissive optical medium.
\newblock {\em Laser \& Photonics Reviews}, 1(4):334--348, 2007.

\bibitem{buckley2020optimization}
Sonia~M Buckley, Alexander~N Tait, Galan Moody, Bryce Primavera, Stephen Olson, Joshua Herman, Kevin~L Silverman, Satyavolu Papa~Rao, Sae Woo~Nam, Richard~P Mirin, et~al.
\newblock Optimization of photoluminescence from w centers in silicon-on-insulator.
\newblock {\em Optics Express}, 28(11):16057--16072, 2020.

\bibitem{tait2020microring}
Alexander~N Tait, Sonia~M Buckley, Jeffrey Chiles, Adam~N McCaughan, S~Olson, S~Papa Rao, SW~Nam, RP~Mirin, and JM~Shainline.
\newblock Microring resonator-coupled photoluminescence from silicon w centers.
\newblock {\em Journal of Physics: Photonics}, 2(4):045001, 2020.

\bibitem{kizhake2024enhanced}
Vijin~Kizhake Veetil, Junyeob Song, Pradeep~N. Namboodiri, Nikki Ebadollahi, Ashish Chanana, Aaron~M. Katzenmeyer, Christian Pederson, Joshua~M. Pomeroy, Jeffrey Chiles, Jeffrey Shainline, Kartik Srinivasan, Marcelo Davanco, and Matthew Pelton.
\newblock Enhanced zero-phonon line emission from an ensemble of w centers in circular and bowtie bragg grating cavities.
\newblock {\em Nanophotonics}, 13(1), 2024.

\bibitem{macquarrie2021generatingTcentres}
ER~MacQuarrie, Camille Chartrand, DB~Higginbottom, KJ~Morse, VA~Karasyuk, Sjoerd Roorda, and Stephanie Simmons.
\newblock Generating t centres in photonic silicon-on-insulator material by ion implantation.
\newblock {\em New Journal of Physics}, 23(10):103008, 2021.

\bibitem{day2024electrical}
Aaron~M Day, Madison Sutula, Jonathan~R Dietz, Alexander Raun, Denis~D Sukachev, Mihir~K Bhaskar, and Evelyn~L Hu.
\newblock Electrical manipulation of telecom color centers in silicon.
\newblock {\em Nature Communications}, 15(1):4722, 2024.

\bibitem{prabhu2023individually}
Mihika Prabhu, Carlos Errando-Herranz, Lorenzo De~Santis, Ian Christen, Changchen Chen, Connor Gerlach, and Dirk Englund.
\newblock Individually addressable and spectrally programmable artificial atoms in silicon photonics.
\newblock {\em Nature Communications}, 14(1):2380, 2023.

\bibitem{islam2023cavity}
Fariba Islam, Chang-Min Lee, Samuel Harper, Mohammad~Habibur Rahaman, Yuqi Zhao, Neelesh~Kumar Vij, and Edo Waks.
\newblock Cavity-enhanced emission from a silicon t center.
\newblock {\em Nano Letters}, 24(1):319--325, 2023.

\bibitem{jhuria2024programmable}
K~Jhuria, V~Ivanov, D~Polley, Y~Zhiyenbayev, W~Liu, A~Persaud, W~Redjem, W~Qarony, P~Parajuli, Qing Ji, et~al.
\newblock Programmable quantum emitter formation in silicon.
\newblock {\em Nature Communications}, 15(1):4497, 2024.

\bibitem{takeda1986precise}
TADAO Takeda, SATOSHI Tazawa, and AKIRA Yoshii.
\newblock Precise ion-implantation analysis including channeling effects.
\newblock {\em IEEE transactions on electron devices}, 33(9):1278--1285, 1986.

\bibitem{johnston2024cavity}
Adam Johnston, Ulises Felix-Rendon, Yu-En Wong, and Songtao Chen.
\newblock Cavity-coupled telecom atomic source in silicon.
\newblock {\em Nature Communications}, 15(1):2350, 2024.

\bibitem{ziegler1995SRIM}
JF~Ziegler.
\newblock The stopping of energetic light ions in elemental matter hzetrn: Description of a free-space ion and nucleon transport and shielding computer program.
\newblock {\em NASA TP-3495}, 1995.

\bibitem{PRXQuantum.5.010102}
Stephanie Simmons.
\newblock Scalable fault-tolerant quantum technologies with silicon color centers.
\newblock {\em PRX Quantum}, 5:010102, Mar 2024.

\bibitem{redjem2023all}
Walid Redjem, Yertay Zhiyenbayev, Wayesh Qarony, Vsevolod Ivanov, Christos Papapanos, Wei Liu, Kaushalya Jhuria, ZY~Al~Balushi, Scott Dhuey, Adam Schwartzberg, et~al.
\newblock All-silicon quantum light source by embedding an atomic emissive center in a nanophotonic cavity.
\newblock {\em Nature communications}, 14(1):3321, 2023.

\bibitem{ebadollahi2024fabrication}
Nikki Ebadollahi, Pradeep~N Namboodiri, Christian Pederson, Vijin~K Veetil, Marcelo~I Davanco, Kartik~A Srinivasan, Aaron~M Katzenmeyer, Matthew Pelton, and Joshua~M Pomeroy.
\newblock Fabrication of silicon w and g center embedded light-emitting diodes for electroluminescence.
\newblock {\em Journal of Vacuum Science \& Technology B}, 42(6), 2024.

\bibitem{buckley2017all}
Sonia Buckley, Jeffrey Chiles, Adam~N McCaughan, Galan Moody, Kevin~L Silverman, Martin~J Stevens, Richard~P Mirin, Sae~Woo Nam, and Jeffrey~M Shainline.
\newblock All-silicon light-emitting diodes waveguide-integrated with superconducting single-photon detectors.
\newblock {\em Applied Physics Letters}, 111(14), 2017.

\bibitem{hollenbach2025programmable}
M~Hollenbach, N~Klingner, P~Mazarov, W~Pilz, A~Nadzeyka, F~Mayer, NV~Abrosimov, L~Bischoff, G~Hlawacek, M~Helm, et~al.
\newblock Programmable activation of quantum emitters in high-purity silicon with focused carbon ion beams.
\newblock {\em Advanced Quantum Technologies}, 8(1):2400184, 2025.

\bibitem{quard2024femtosecond}
Hugo Quard, Mario Khoury, Andong Wang, Tobias Herzig, Jan Meijer, S{\'e}bastien Pezzagna, S{\'e}bastien Cueff, David Grojo, Marco Abbarchi, Hai~Son Nguyen, et~al.
\newblock Femtosecond-laser-induced creation of g and w color centers in silicon-on-insulator substrates.
\newblock {\em Physical Review Applied}, 21(4):044014, 2024.

\bibitem{gu2025end}
Qiushi Gu, Valeria Saggio, Camille Papon, Alessandro Buzzi, Ian Christen, Christopher Panuski, Carlos Errando-Herranz, and Dirk Englund.
\newblock End-to-end physics-based modeling of laser-activated color centers in silicon.
\newblock {\em arXiv preprint arXiv:2501.17240}, 2025.

\end{thebibliography}
\bibliographystyle{unsrt} 

\end{document}